\documentclass[%
aip,
jmp,%
amsmath,amssymb,
reprint,%
]{revtex4-1}

\usepackage{graphicx}
\usepackage{dcolumn}
\usepackage{bm}
\usepackage{subcaption}
\usepackage{subfig}
\usepackage{cleveref}
\begin{document}
\preprint{AIP/123-QED}
\title[]{Quantum photonics hybrid integration platform}

\author{E. Murray}
\affiliation{ Cambridge Research Laboratory, Toshiba Research Europe Limited, 208 Science Park, Milton Road, Cambridge, CB4 0GZ, United Kingdom. }
\affiliation{ 
Cavendish Laboratory, University of Cambridge, J.J. Thomson Avenue, Cambridge CB3 0HE, United Kingdom.
}

\author{D. J. P. Ellis}
\affiliation{ 
Cambridge Research Laboratory, Toshiba Research Europe Limited, 208 Science Park, Milton Road, Cambridge, CB4 0GZ, United Kingdom.
}

\author{T. Meany}
\affiliation{ Cambridge Research Laboratory, Toshiba Research Europe Limited, 208 Science Park, Milton Road, Cambridge, CB4 0GZ, United Kingdom. }

\author{F. F. Fl\"other}
\affiliation{ Cambridge Research Laboratory, Toshiba Research Europe Limited, 208 Science Park, Milton Road, Cambridge, CB4 0GZ, United Kingdom. }
\affiliation{ 
Cavendish Laboratory, University of Cambridge, J.J. Thomson Avenue, Cambridge CB3 0HE, United Kingdom.
}

\author{J. P. Lee}
\affiliation{ Cambridge Research Laboratory, Toshiba Research Europe Limited, 208 Science Park, Milton Road, Cambridge, CB4 0GZ, United Kingdom. }
\affiliation{ 
Engineering Department, University of Cambridge, 9 J. J. Thomson Avenue, Cambridge, CB3 0FA, United Kingdom.
}

\author{J. P. Griffiths}
\affiliation{ 
Cavendish Laboratory, University of Cambridge, J.J. Thomson Avenue, Cambridge CB3 0HE, United Kingdom.
}

\author{G. A. C. Jones}
\affiliation{ 
Cavendish Laboratory, University of Cambridge, J.J. Thomson Avenue, Cambridge CB3 0HE, United Kingdom.
}

\author{I. Farrer}
\affiliation{ 
Cavendish Laboratory, University of Cambridge, J.J. Thomson Avenue, Cambridge CB3 0HE, United Kingdom. 
}

\author{D. A. Ritchie}
\affiliation{ 
Cavendish Laboratory, University of Cambridge, J.J. Thomson Avenue, Cambridge CB3 0HE, United Kingdom.
}

\author{A. J. Bennett}
\email{anthony.bennet@crl.toshiba.co.uk}
\affiliation{ 
Cambridge Research Laboratory, Toshiba Research Europe Limited, 208 Science Park, Milton Road, Cambridge, CB4 0GZ, United Kingdom.
}

\author{A. J. Shields}
\affiliation{ 
Cambridge Research Laboratory, Toshiba Research Europe Limited, 208 Science Park, Milton Road, Cambridge, CB4 0GZ, United Kingdom.
}

\begin{abstract}
Fundamental to integrated photonic quantum computing is an on-chip method for routing and modulating quantum light emission. We demonstrate a hybrid integration platform consisting of arbitrarily designed waveguide circuits and single photon sources. InAs quantum dots (QD) embedded in GaAs are bonded to an SiON waveguide chip such that the QD emission is coupled to the waveguide mode. The waveguides are SiON core embedded in a SiO$_2$ cladding. A tuneable Mach Zehnder modulates the emission between two output ports and can act as a path-encoded qubit preparation device. The single photon nature of the emission was verified by an on-chip Hanbury Brown and Twiss measurement.
\end{abstract}

\maketitle

Linear optical quantum computing has been proven to be computationally efficient with single photon sources and a series of beamsplitters and phase shifters \cite{knill2001scheme}. Although few photon gates have been demonstrated using bulk optics \cite{o2003demonstration}, scaling to more complex circuits requires integrated photonic technology \cite{carolan2015universal}.

Integrated photonics offers the potential for true scalability due to component miniaturisation. Stability is intrinsic to the platform and offers a reduction in complexity and size of the device \cite{politi2009integrated}. Many of the elements needed for linear optical quantum computing can be manufactured on-chip. High fidelity beam splitters and Mach Zehnders (MZs) can be made with various semiconductors platforms \cite{wang2014gallium, zhang2011, politi2008silica} as well as on-chip detectors \cite{gerrits2011chip, hadfield2009single}. However, thus far integration of single photon sources into low loss waveguides remains and open issue.

Semiconductor III-V QDs have been shown to produce bright, single photon emission \cite{Bennett:05}, emit indistinguishable and entangled photons \cite{he2013demand,stevenson2012}, can be site-controlled \cite{juska2013towards} and compatible with semiconductor foundry techniques. Various approaches for embedding QDs into integrated circuits are being explored. Photonic crystal waveguides yield high coupling efficiency of the QD emission into the in-plane propagating waveguide mode \cite{schwagmann2011chip, arcari2014near}. They can produce in-plane indistinguishable photons \cite{kalliakos2014plane}, however as of yet no directional couplers or active modulators with embedded QDs have been demonstrated.  QDs embedded in ridge waveguides in GaAs have been reported combined with on-chip superconducting single photon detectors \cite{reithmaier2013chip}. Air clad GaAs ridge waveguides have also demonstrated QD integrated directional couplers \cite{prtljaga2014, jons2014monolithic}. Other approaches use heralded single photons from spontaneous parametric down conversion integrated with waveguide chips \cite{meany2014hybrid} however this approach lacks deterministic emission.

In this letter we present a novel platform for hybrid integration of III-V QDs with silicon oxynitride waveguides. A GaAs chip containing InAs quantum dots are bonded orthogonally to the SiON chip such that the photons emitted from the surface of the GaAs chip are routed into a guided mode. The quantum dots are embedded in a distributed Bragg reflector cavity with alternating layers of AlAs and GaAs. The SiON chip consists of a waveguide to deliver laser excitation light to the quantum dots and a return line consisting of a MZ interferometer. The MZ can act as a qubit preparation device for the single photons emitted from the QD. If a single photon impinges on the MZ it will be placed into a path encoded superposition of each output mode. The probability amplitude of being in each mode is chosen by the tuning of the MZ. This SiON technology is compatible with the creation of arbitrary designs of beamsplitters, MZs and phase shifters.

Orthogonal bonding allows the surface emission from the QD to be optimised by growing a distributed Bragg reflector cavity and/or creating a micropillar structure. Previous reports show efficiencies of out of plane QD collection as high as 0.75 into free space high numerical aperture objectives \cite{claudon2010highly, munsch2013dielectric}. In SiON the waveguide numerical aperture is 0.3, given by $\sqrt{n_{core}^2 - n_{clad}^2}$, where $n_{core} = \mathrm{1.55}$ and $n_{clad} = \mathrm{1.51}$. The hybrid platform also has the potential for diode structures to be created for electrically driven or tuneable QD devices integrated with the waveguides. 


\begin{figure}[h!]
\begin{center}
\begin{subfigure}{1\linewidth}
\includegraphics[width=0.9\linewidth]{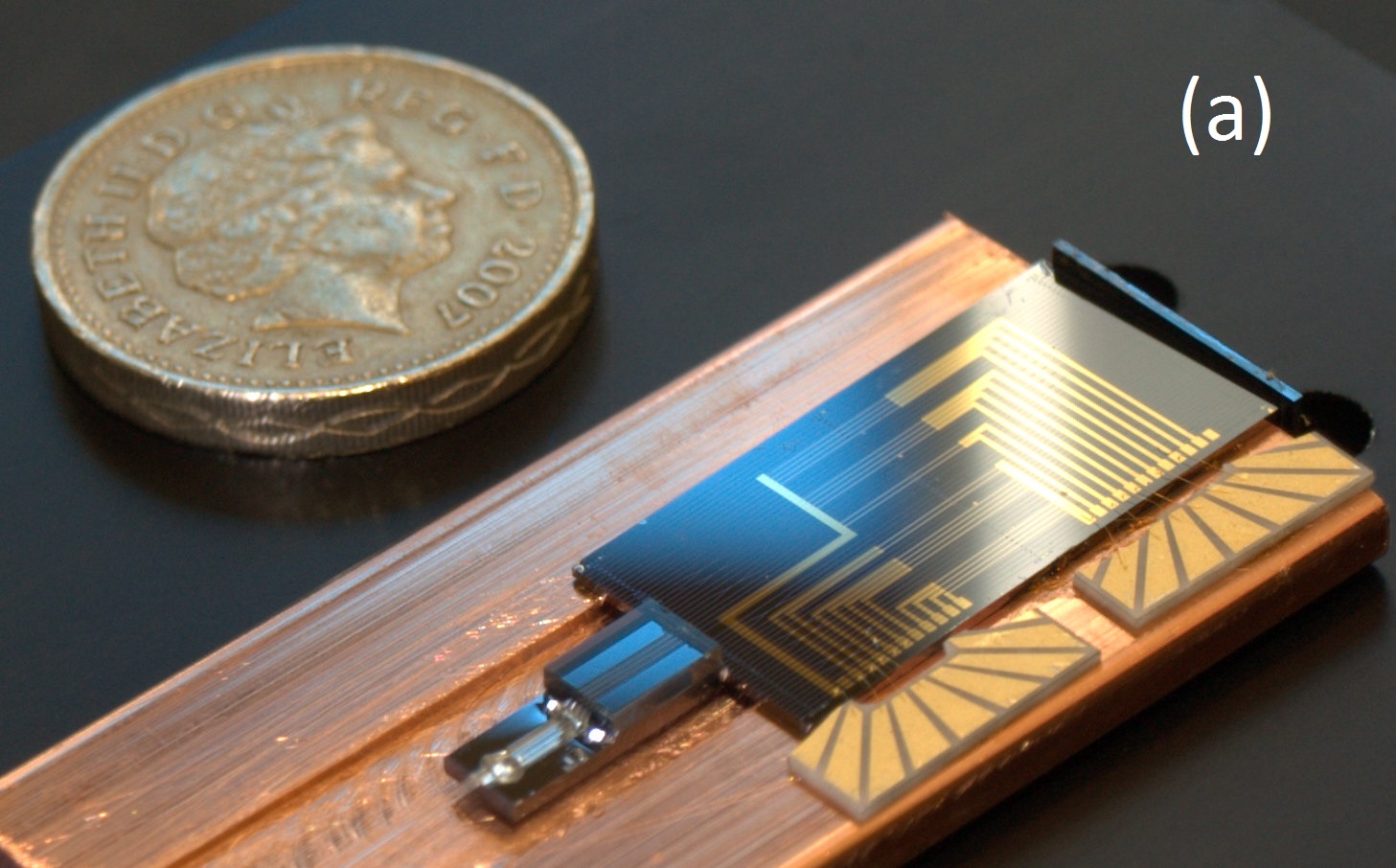}
\label{fig:photo}
\end{subfigure}\\[2ex]

\begin{subfigure}{1\linewidth}
\includegraphics[width=1\linewidth]{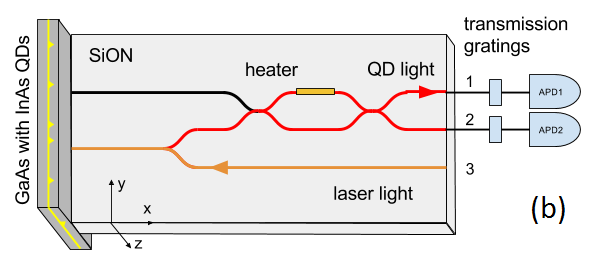}
\label{fig:schematic}
\end{subfigure}\\[2ex]
\begin{subfigure}{1\linewidth}
\includegraphics[width=0.95\linewidth]{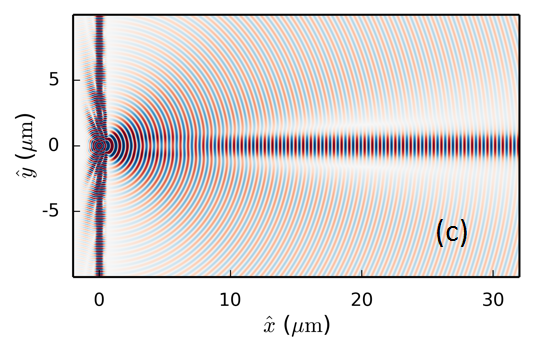}
\label{fig:sim}
\end{subfigure}
\end{center}
\caption{(a) Photograph of a bonded device. (b) Schematic of the prototype device. The orange line indicates the waveguide to deliver excitation light; the red line indicates the waveguides travelled by the QD emission. (c) Finite difference time domain simulation of a QD dipole emitting in a cavity and the emission being guided by the SiON core. }
\label{fig:images}
\end{figure}

Figure \ref{fig:images} (a) shows a photograph of a hybrid device. A strip of III-V containing QDs is bonded to one end of the SiON waveguide chip. The QD source was grown by molecular beam epitaxy. A QD density was chosen such that a dot which is emits in the centre cavity mode is aligned with a waveguide. The central cavity wavelength is at 910nm. The SiON device was fabricated by plasma enhanced chemical vapour deposition to define a layer of SiO$_2$ undercladding and SiON core on a silicon substrate. Electron-beam lithography and reactive ion etching were used to define the SiON core profile before finally an SiO$_2$ overclad layer was deposited. The orthogonal bonding method allows the surface emission from the QD to be collected by the waveguide. The photons are routed into the waveguides, two sequential directional couplers form a MZ with a nickel-chromium alloy heater applying a local phase shift to one MZ arm.  A polarisation maintaining V-groove array is aligned and attached to the waveguides for collection. The device is kept at 4K for the duration of the experiments. 

Figure \ref{fig:images} (b) shows the optical schematic of the experiments. A single channel delivers laser light. The QD light is returned through the MZ and collected into fibre by the V-groove array. The V-groove fibres are sent directly to a spectrometer. For time resolved experiments transmission gratings are used to spectrally filter the emission before sending the light to avalanche photodiodes.

The characteristics of the device were simulated by using the finite difference time domain package MEEP \cite{oskooi2010meep, mandelshtam1997harmonic}. A $\hat{z}$ oriented dipole emitter was placed in the centre of the cavity spacer aligned to the centre of the waveguide. A perfectly matched layer was placed at the edges of the simulation domain to absorb all light and prevent unwanted reflections. The $\hat{z}$ component of the electric field is shown in Figure \ref{fig:images} (c). There is a clear emission pattern along the waveguide core. 

\begin{figure}[h!]
\begin{subfigure}{0.60\linewidth}
\includegraphics[width=1\linewidth]{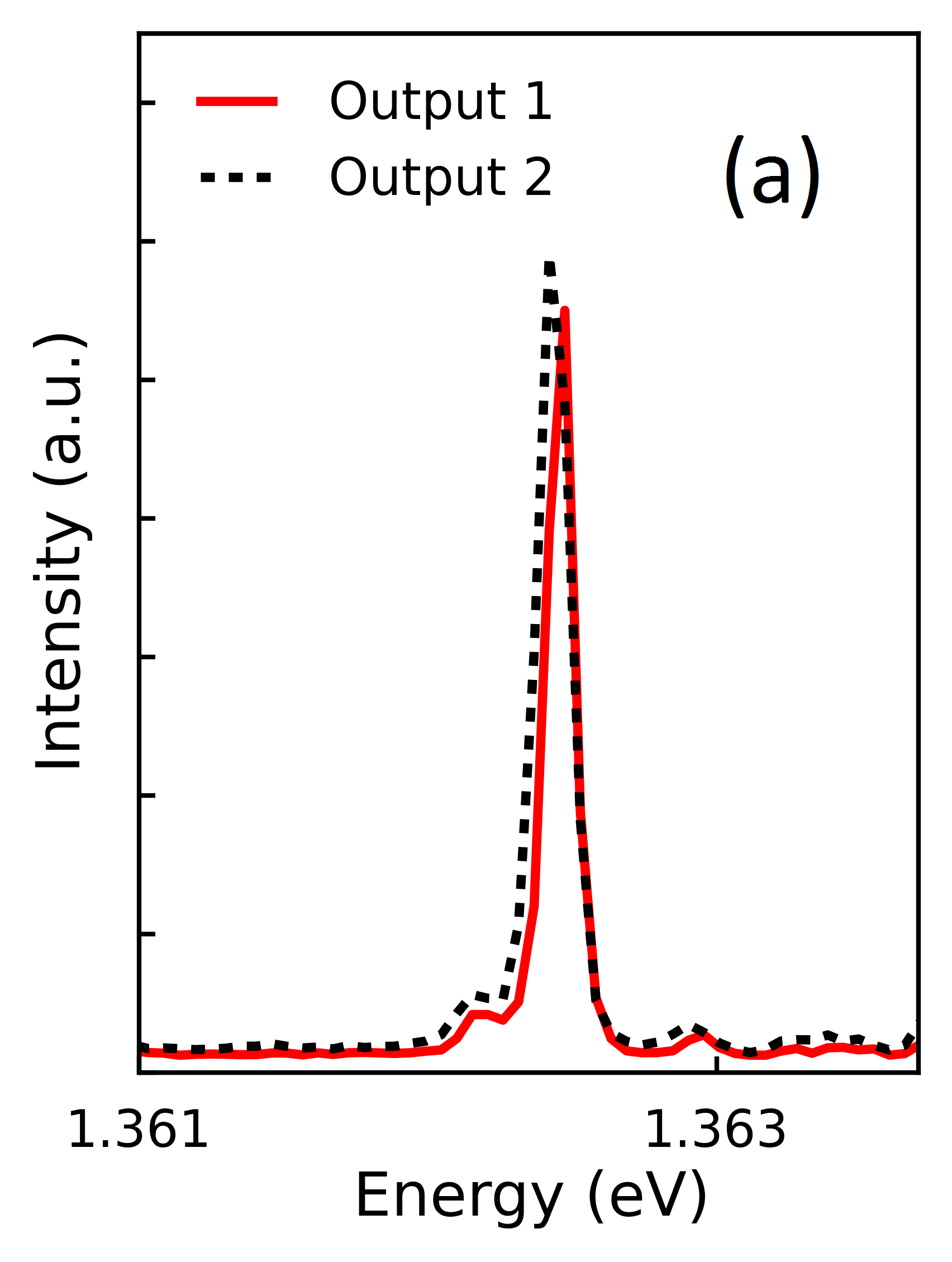}
\end{subfigure}
\begin{subfigure}{0.38\linewidth}

\begin{subfigure}{0.99\linewidth}
\includegraphics[width=1\linewidth]{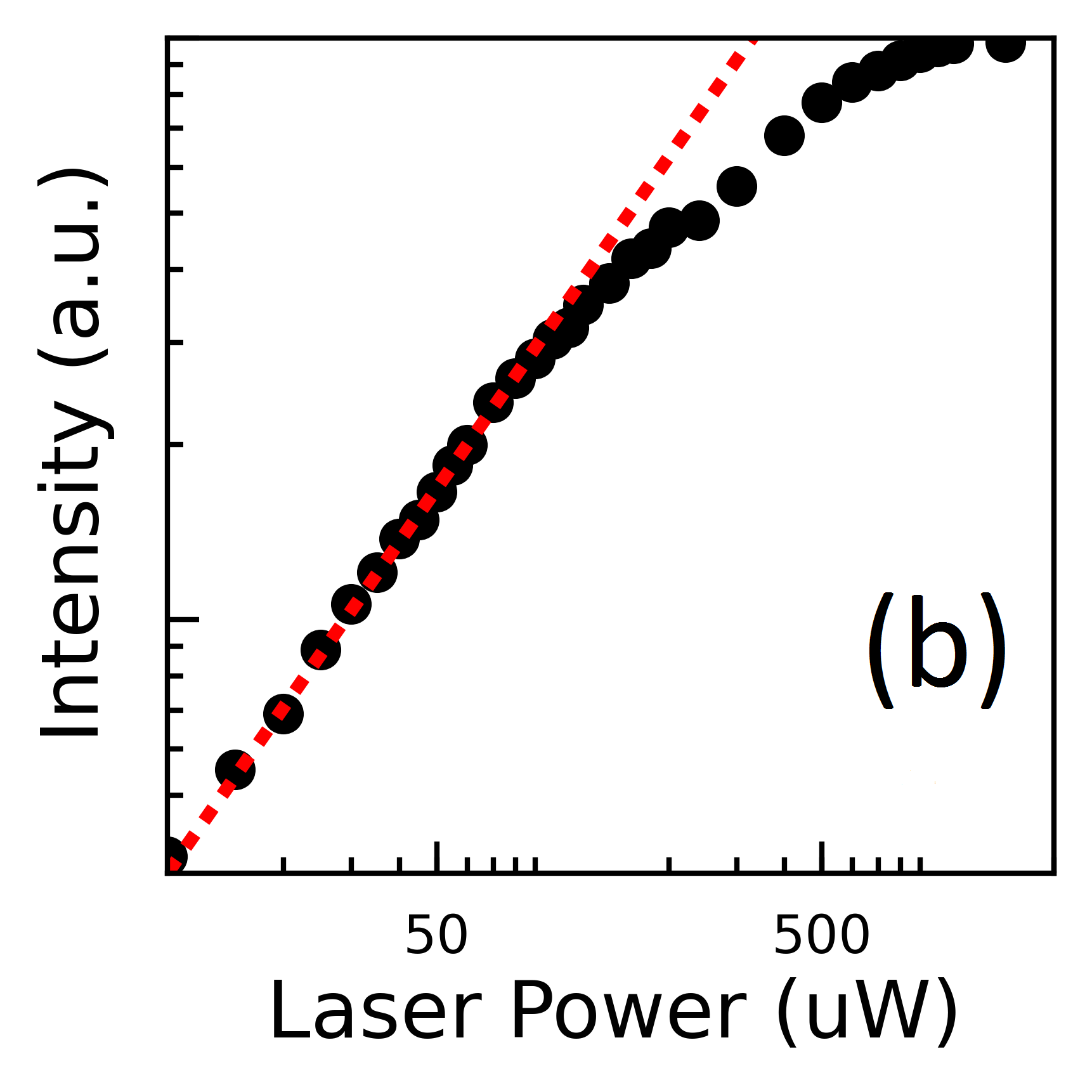}
\end{subfigure}\\[2ex]
\begin{subfigure}{0.99\linewidth}
\includegraphics[width=1\linewidth]{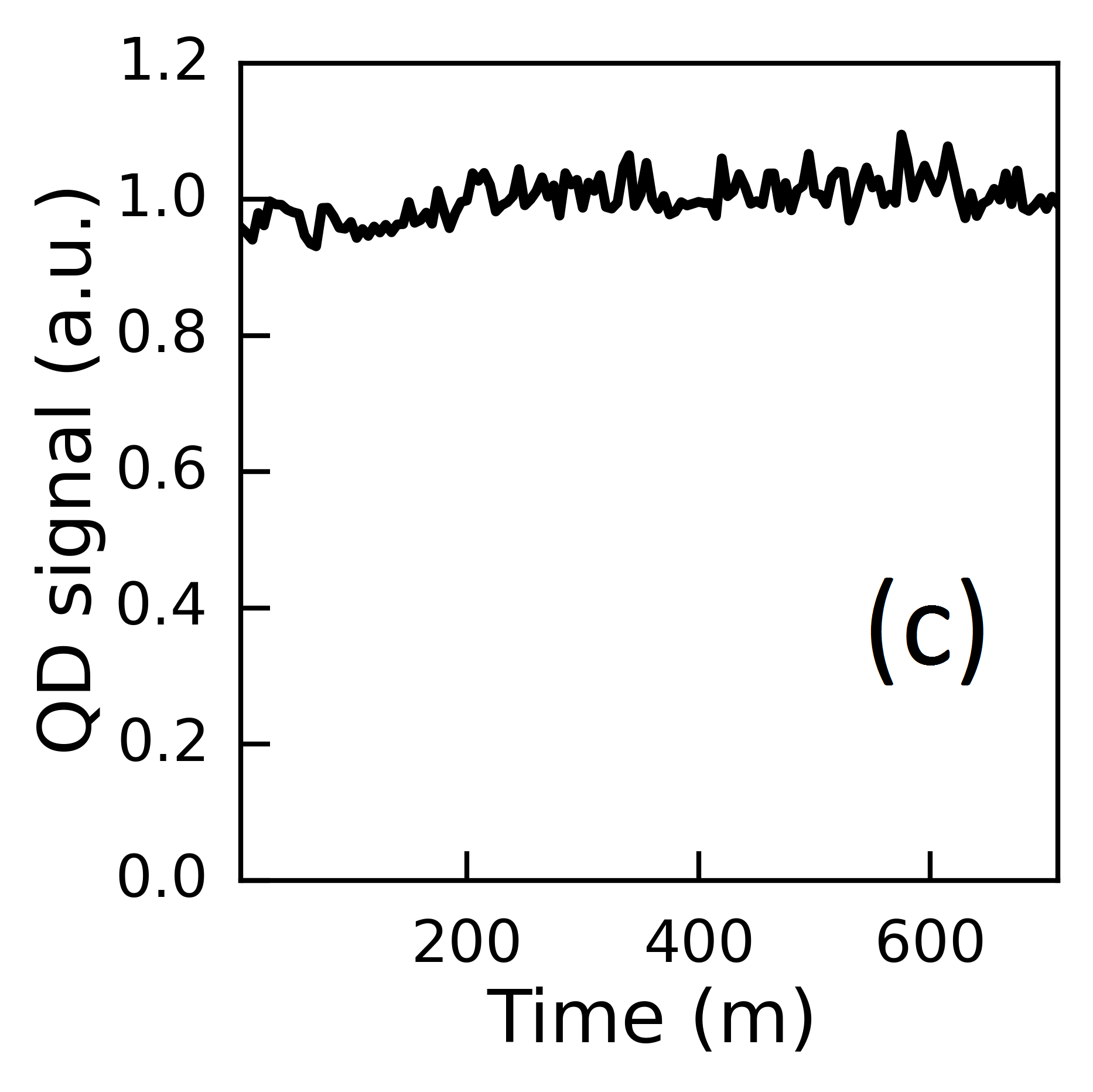}
\end{subfigure}

\end{subfigure}
\caption{(a) Spectrum from both output ports of the device indicated in red and black. (b) Power dependence of the spectral line at 1.362 eV. (c) Power output from the QD line over the course over 12 hours.}
\label{fig:spec}
\end{figure}

The efficiency of this device was determined theoretically. A bounding box which records the Fourier transformed fields was placed at the edge of the domain and just inside the perfectly matched layer. From this bounding box the total power spectrum is recorded when the system is excited with a short Gaussian pulse. Another flux plane is placed across the waveguide. The waveguide core size was 1.6 $\mu \mathrm{m}$ and the far field propogating mode has a spatial $1/e$ width of 1.88 $\mu \mathrm{m}$ which was chosen as the size of the waveguide flux plane. It is placed sufficiently far from the surface of the III-V so that only the waveguide propagating mode is measured. Taking the ratio of the light propagating in the waveguide to the total in the bounding box gives the efficiency of the QD emission into the waveguide to be 2.8\%. This is considerably higher than for a QD in bulk of 0.5\% and could be further enhanced with higher Q cavities.



\begin{figure}[h!]
\begin{subfigure}{0.49\linewidth}
\includegraphics[width=1\linewidth]{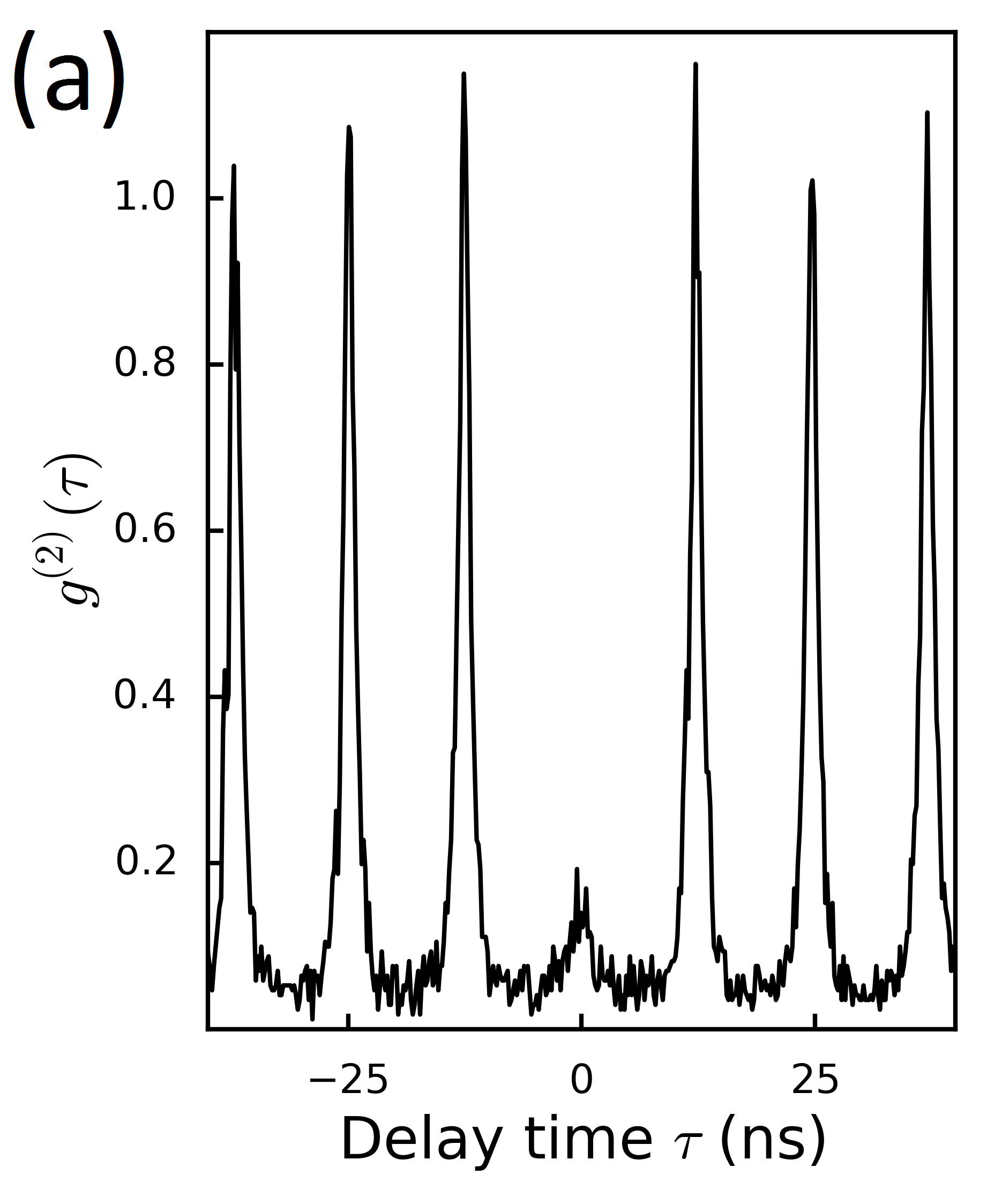}
\label{fig:g2p}
\end{subfigure}
\begin{subfigure}{0.49\linewidth}
\includegraphics[width=1\linewidth]{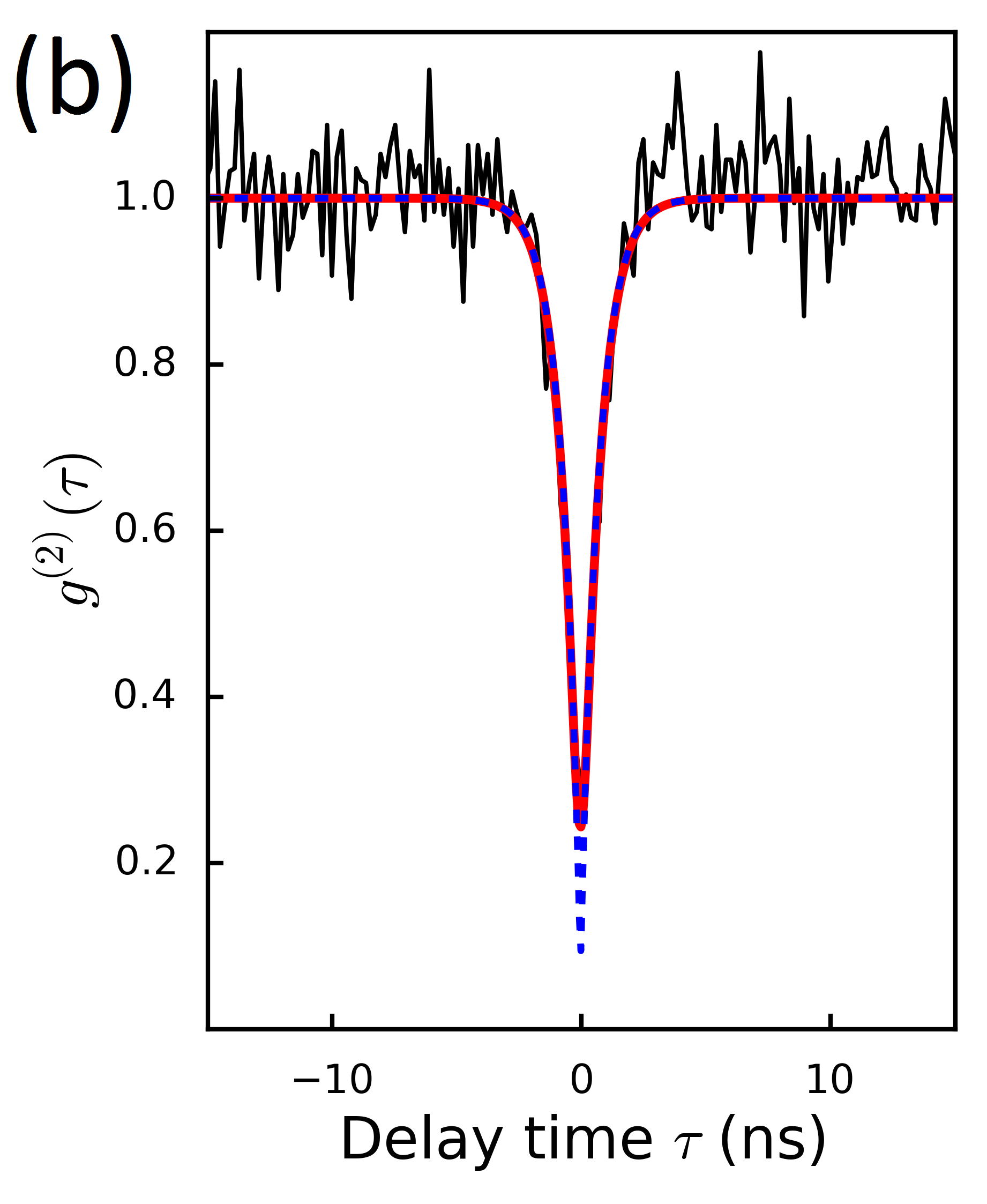}
\label{fig:g2cw}
\end{subfigure}
\caption{(a) Second order correlation function under pulsed excitation (b) Second order correlation function under continuous wave excitation. The black line is the raw data, the red line is a fit taking the system response funtion into account, and the dashed blue line is the fit deconvoluted with the response function.}
\label{fig:g2s}
\end{figure}

The integrated device is fixed once the bonding is complete. This creates stability and no drifting in emission intensity is seen over the course of 12 hours. The QD intensity as a function of time is plotted in Figure \ref{fig:spec} (c). The MZ was tuned to 50:50 and the spectrum from both outputs of the device can be seen in Figure \ref{fig:spec} (a). The QD analysed was located at the centre of the cavity mode at 1.362 eV. A power dependent spectum was recorded as seen in Figure \ref{fig:spec} (b). The emission, before saturation, is approximately linear (P$^{0.95}$) implying an exciton and not a higher order complex \cite{grundmann1997theory}. The peak does not exhibit the polarisation splitting which is characteristic of a neutral exciton\cite{bayer2002fine,gammon1996fine} and is therefore likely to derive from a charged exciton.

To verify the single photon nature of the QD emission, a Hanbury Brown and Twiss experiment was recorded using the on-chip MZ as a beamsplitter. The emission from both arms was sent through two different transmission gratings for spectral filtering and then sent to avalanche photodiodes. An absence of correlation coincidences at time $\tau = 0$ implies a single quantum state is under study. The second order correlation curves were taken under continuous wave ($\lambda$ = 810nm) and pulsed ($\lambda$ = 850nm) excitation. The curves are shown in Figure \ref{fig:g2s}. In the case of the pulsed curve (shown in Figure \ref{fig:g2s} (a)) the signal to noise ratio due to dark counts of the detectors was calculated and subtracted. A time window was also applied to the data, since the lifetime of the QD was 670 $\pm$ 3ps, the vast majority (96\%) of QD emission resides in a 4ns window. So only coincidences inside this window were used for calculating the $g^{(2)}(0)$ value of 0.19.

In the CW case as seen in Figure \ref{fig:g2s} (b) the black line corresponds to the data, the red solid line corresponds to a fit and the blue line to a deconvoluted fit. The deconvolution was done to subtract the response function of the detectors and timing system. The $g^{(2)}(0)$ was taken from the deconvoluted fit to be 0.09.

\begin{figure}[h!]
\centering
\includegraphics[width=0.9\linewidth]{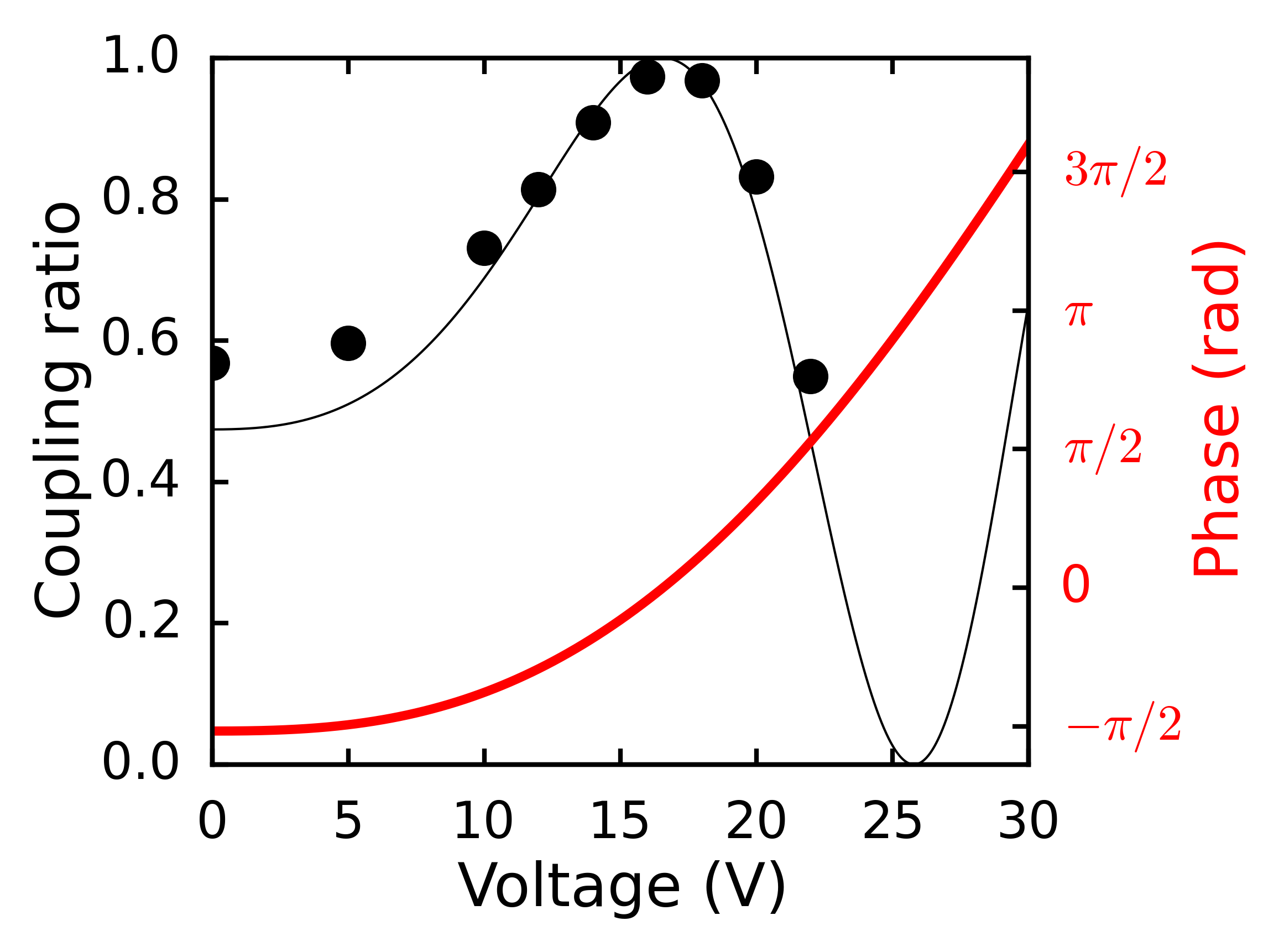}
\caption{The measured coupling ratio of the MZ was a function of applied voltage, the solid black line is a fit to the data. The red line is the calculated phase.}
\label{fig:modul}
\end{figure}

The active modulation of the MZ was tested on the single photon source. A voltage was applied to the heater on one arm of the MZ which induces local heating of the arm and a change in the refractive index of the waveguide core. This creates a relative phase between the light in each arm. The emission coupling to each output arm of the MZ then varies as a function of the applied voltage. The coupling ratio of the device is defined as the power difference in each arm divided by the sum. The coupling ratio along with calculated phase\cite{matthews2009manipulation}, is shown in Figure \ref{fig:modul}.

In conclusion we have demonstrated a novel method for integrating a III-V quantum light source with an SiON waveguide platform. The quantum nature of the source was verified using on-chip components and the active modulation of the emission was demonstrated. This device shows potential for integration of site controlled \cite{jamil2014chip, juska2013towards} QDs granting precise alignment of multiple QDs with multiple waveguides allowing for scalable quantum manipulation. 

\vspace{1cm}

EM and TM acknowledge support by the Marie Curie Actions within the Seventh Framework Programme for Research of the European Commission, under the Initial Training Network PICQUE, Grant No. 608062.
FF acknowledges support from both the EPSRC and Toshiba Research Europe Ltd in Cambridge. JL acknowledges support from both the EPSRC CDT in Photonic Systems Development and Toshiba Research Europe Ltd in Cambridge. The authors acknowledge funding from the EPSRC for MBE system used in the production of the QD samples. 

\bibliographystyle{apl}

\end{document}